\documentclass[runningheads]{llncs}
\usepackage[T1]{fontenc}
\usepackage{appendix}
\usepackage{amsmath}
\usepackage{amssymb}
\usepackage{float}
\usepackage{svg}
\usepackage{array}      
\usepackage{caption}
\usepackage{subcaption}
\usepackage{booktabs} 
\usepackage{mathtools} 
\usepackage{marvosym} 
\usepackage{graphicx}
\begin{document}
\title{Cognitive Effort in the Two-Step Task:  An Active Inference Drift-Diffusion Model Approach}

\titlerunning{Cognitive Effort in the Two-Step Task}

\author{\'Alvaro Garrido-P\'erez\inst{1}\orcidID{0009-0003-5481-8166} \and
Viktor Lemoine\inst{1}\and
Amrapali Pednekar \inst{1}\orcidID{0009-0005-6194-3955} \and
Yara Khaluf \inst{2}\orcidID{0000-0002-5590-9321} \and
Pieter Simoens \inst{1}\orcidID{0000-0002-9569-9373}}
\authorrunning{A. Garrido-P\'erez et al.}
\institute{IDLab, Department of Information Technology, Ghent University - imec, Belgium \\
\email{alvaro.garridoperez@ugent.be}
\and
Wageningen University \& Research, Wageningen, The Netherlands}

\maketitle              % typeset the header of the contribution
\begin{abstract}

High-level theories rooted in the Bayesian Brain Hypothesis often frame cognitive effort as the cost of resolving the conflict between habits and optimal policies. In parallel, evidence accumulator models (EAMs) provide a mechanistic account of how effort arises from competition between the subjective values of available options. Although EAMs have been combined with frameworks like Reinforcement Learning to bridge the gap between high-level theories and process-level mechanisms, relatively less attention has been paid to their implications for a unified notion of cognitive effort. Here, we combine Active Inference (AIF) with the Drift-Diffusion Model (DDM) to investigate whether the resulting AIF-DDM can simultaneously capture the effort arising from both habit violation and value discriminability. To our knowledge, this is the first time AIF has been combined with an EAM. We tested the AIF-DDM on a behavioral dataset from the two-step task and compared its predictions to an information-theoretic definition of cognitive effort based on AIF. The model's predictions successfully accounted for second-stage reaction times but failed to capture the dynamics of the first stage. We argue the latter discrepancy likely stems from the experimental design rather than a fundamental flaw in the model's assumptions about cognitive effort. Accordingly, we propose several modifications of the two-step task to better measure and isolate cognitive effort. Finally, we found that integrating the DDM significantly improved parameter recovery, which could help future studies to obtain more reliable parameter estimates.

\vspace{1em}

\end{abstract}

\section{Introduction}
Building upon the Bayesian Brain Hypothesis (BBH), numerous studies in the past decade have tried to formalize cognitive effort using information-theoretic principles \cite{zenon2019information}. According to the BBH, humans maintain an internal world model encoded in prior beliefs, which they continuously update as they interact with the environment. Within this context, cognitive effort arises from the conflict between a pre-existing belief about how to act (a habit) and an updated belief about the optimal policy \cite{parr2023cognitive}.

In a decision-making task, cognitive effort may also arise from the competition between the subjective values of the available choices. When these values are closer together—that is, when value discriminability is low—reaction times (RTs) tend to increase (e.g., \cite{lee2023risky,fiedler2012dynamics}). Although RTs are not a direct measure of cognitive effort, they are often used as a proxy (e.g., \cite{yao2025neural}), based on the assumption that slower responses reflect more information processing.

From an information-theoretic perspective, the effect of value discriminability on RTs can be understood as the additional cognitive effort required to resolve increased choice uncertainty \cite{proctor2018hick,mcdougle2021modeling,lai2024human}. Yet, information theory provides no explicit account of the underlying deliberation process, which complicates the task of linking its predictions to specific neural signatures. In contrast, evidence accumulator models (EAMs) offer a mechanistic explanation for how value competition shapes decision speed, which may be empirically tested (e.g., \cite{steinemann2024direct}).

%However, cognitive effort depends not only on the divergence between prior and posterior beliefs, but also on the competition between the subjective values of the available options. Specifically, effort increases as the difference between these subjective values—or value discriminability—decreases. From an information-theoretic perspective, this can be framed as the effort required to resolve uncertainty in the prior belief about which action to take \cite{zenon2019information}. Yet, information theory provides no explicit description of the underlying deliberation process, making it challenging to link its predictions to specific neural signatures. In contrast, evidence accumulator models (EAMs) provide a powerful mechanistic explanation for how value competition leads to longer reaction times (RTs) \cite{steinemann2024direct}. Although RTs are not a direct measure of cognitive effort, tasks with greater choice uncertainty- requiring more information processing - result in longer deliberation times \cite{proctor2018hick,mcdougle2021modeling,lai2024human}. Thus, RTs are frequently employed as a practical proxy for cognitive effort (e.g., \cite{yao2025neural}).

A major limitation of using EAMs to study cognitive effort is that they are agnostic to how beliefs are formed. This limitation is often addressed by combining EAMs with Reinforcement Learning (RL) \cite{pedersen2017drift}. However, in recent years, Active Inference (AIF) has emerged as a powerful alternative to RL, offering a first-principles perspective on perception, learning, and decision-making \cite{hodson2024empirical}.

Here, we investigate whether integrating a Drift Diffusion Model (DDM), a prominent class of EAM, with AIF can simultaneously capture the influence of both value discriminability and habit violation on cognitive effort. To our knowledge, this is the first attempt to combine AIF with an EAM to model human behaviour. We evaluate the integrated AIF-DDM using the two-step task, a version of the multi-armed bandit in which participants must plan two steps ahead \cite{daw2011model}. Furthermore, we compare its predictions with a recently proposed definition of cognitive effort in AIF \cite{parr2023cognitive}. 

Our work builds directly on a recent study that developed an AIF model of the two-step task \cite{gijsen2022active}. The study demonstrated that AIF outperformed a Hybrid Reinforcement Learning (HRL) model (which combines model-free and model-based strategies \cite{shahar2019improving}) in two out of four datasets, while achieving comparable performance in the remaining two. Moreover, the authors provided compelling evidence for directed exploration—a key differentiator between AIF and HRL. Despite these achievements, the study could not determine which specific AIF learning mechanisms participants were using. Given prior evidence that integrating a DDM into an HRL model improved the reliability of model-based estimates \cite{shahar2019improving}, we tested whether the combined AIF–DDM could resolve this ambiguity.

\section{Methods}

\subsection{Participants and behavioural task}\label{methods:pars_and_behavioural_task}
In this section, we will briefly describe the two-step task and the behavioural dataset that we used to fit the computational models. For more details on the experimental procedure, we encourage reading the paper that made the dataset publicly available \cite{feher2020humans}.

\begin{figure}
    \centering
    %\includesvg[width=0.9\textwidth]{Figures/two_step_task.drawio.svg}
    %\includegraphics[width=0.9\textwidth]{Figures/two_step_task.drawio.png}

    \includegraphics[width=0.9\textwidth]{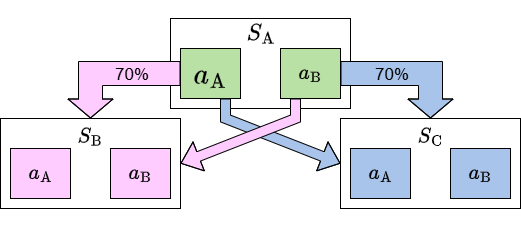}

    \caption{Abstract representation of the two-step task. At the first stage (top), a choice leads to one of two second-stage states (bottom). Transitions are either common (p=0.7, thick arrows) or rare (p=0.3). The two initial actions have opposing common transitions. A second-stage choice may result in a monetary reward with a probability that fluctuates over time.}
    \label{fig:abstract_rep_two_step_task}
\end{figure}

As the name suggests, each trial of the two-step task consists of two stages (Fig. \ref{fig:abstract_rep_two_step_task}). In the first stage, participants choose between two actions. Each action leads to one of the two second-stage states through a probabilistic transition that is either common ($p = 0.7$) or rare ($p = 0.3$). Importantly, the two first-stage actions have opposite most-likely transitions. These transition probabilities remain constant throughout the task. After transitioning to a second-stage state, participants make a final choice between two actions. Each of these second-stage actions results in a monetary reward or no reward, depending on its current \textit{outcome} probability. In contrast to the fixed transitions, these outcome probabilities fluctuate independently over time, following Gaussian random walks.

%\vspace{-0.3cm}

%We analyzed data from the "Magic Carpet" dataset \cite{feher2020humans}, which initially included 24 participants. In the experiment, participants had a 2-second deadline to respond at each stage. We first preprocessed the data by excluding invalid trials, defined as those with RTs under 100 ms or those that missed the 2-second deadline. We then applied a participant-level exclusion criterion, removing any individual for whom more than 10\% of trials were invalid.
%One participant exceeded this threshold, with 44.1\% of trials being either missed or too fast (RT < 100 ms). This participant was excluded, resulting in a final sample of n=23. In total, 6.53\% of all trials from the original dataset were excluded from the analysis, including all data from the outlier.

We analysed data from the "Magic Carpet" dataset, which was made available by \cite{feher2020humans} and originally comprised 24 participants. In this experiment, participants had a 2-second deadline to respond at each stage. Every time this deadline was surpassed, the trial was labelled as a
\textit{missed} trial and the participant moved on to the next stage or trial. Before conducting any analysis, we identified and removed these trials. We also considered invalid trials those with at least one RT smaller than 100 ms (to exclude anticipatory responses that are too fast to reflect genuine deliberation \cite{myers2022practical}). Finally, any participant with more than 10\% invalid trials was removed from the dataset. One participant was excluded under this criterion, with 44.1\% of their trials being either missed or too fast. Therefore, the original sample size was reduced to n=23. In total, 6.53\% of the data was excluded from the analysis (including all the data from the removed participant).

\subsection{An active inference drift-diffusion model of the two-step task}\label{sec:aif_ddm}
In this section, we introduce the AIF model developed by \cite{gijsen2022active} and discuss how we integrate it with a DDM. Further explanations of the equations are provided in Appendix \ref{apx_aif_learning_rules}. To establish a comparative benchmark, we implemented an HRL-DDM. Since the latter model is secondary to our main analysis, its mathematical formulation is only detailed in Appendix \ref{apx:hrl_ddm}.

In what follows, we will use the same notation as in \cite{gijsen2022active}. For a given trial $t$, the first-stage state and chosen actions are denoted by $s_{1,t}$ and $a_{1,t}$, respectively. Likewise, the second-stage state and chosen action are denoted by $s_{2,t}$ and  $a_{2,t}$, respectively. Note that the first-stage state can only be $s_{\text{A}}$, but the second-stage state may be $s_{\text{B}}$ or $s_{\text{C}}$, depending on the transition (see Fig. \ref{fig:abstract_rep_two_step_task}). Outcomes $o_t \in \{0, 1\}$ represent the observed reward (or absence of it) after choosing a second-stage action.

According to the AIF model developed by \cite{gijsen2022active}, an agent performing the two-step task is equipped with a generative model (i.e., a set of beliefs about how the task environment evolves given its actions). In the two-step task, agents must learn the outcome probabilities of each of the four second-stage actions. We will refer to the set of these four probabilities as $\theta$, and the belief distribution over $\theta$, at a given trial t, $\pi_{t}(\theta)$. In addition, the agent must learn the four transition probabilities $p(s_{2,t}|s_{1,t}, a_{1,t})$. However, as pointed out by \cite{gijsen2022active}, these transitions are accurately learned in a few trials and therefore, action-selection is only sensitive to information regarding outcome probabilities.

Thus, an agent performing the two-step task must optimally balance getting as many rewards as possible (exploitation) and learning the four outcome probabilities (exploration). According to AIF, this balance is achieved by selecting actions that minimize a quantity known as Expected Free Energy (EFE), which in this case is given by the following equation:

\begin{equation}
G_t(a) = \underbrace{- \mathbb{E}_{p(o_t; \pi_t(\theta)|a)} [\ln p(o_t|C)]}_{\text{\shortstack{Realising preferences \\ (Exploitation)}}} - \underbrace{\mathbb{E}_{p(o_t; \pi_t(\theta)|a)} [D_{\text{KL}}(\pi_t(\theta)|o_t, a \| \pi_t(\theta))]}_{\text{\shortstack{Model parameter exploration \\ (Active Learning)}}}
\label{eq:EFE_definition} 
\end{equation}

Where $G_t(a)$ is the EFE of a given action at a given trial, $p(o_{t}|C)$ is the distribution over prior preferred observations (Eq. \ref{eq:preferred_obs}, Appendix \ref{apx_aif_learning_rules}), which depends on a free parameter $\lambda$. Note that in the two-step task, the \textit{preferred} observation is to get the reward after a second-stage choice. $D_{KL}$ is the Kullback-Leibler divergence between prior beliefs about second-stage outcome probabilities $\pi_{t}(\theta)$ and posterior beliefs after selecting an action and observing its outcome. $p(o_t;\pi_{t}(\theta)|a)$ is a distribution equal to $p(o_t|\theta, a)\pi_t(\theta)$. Note that, unlike traditional EFE formulations, there is no \textit{hidden state exploration} term because in the real experiment participants can always see the state in which they are \cite{schwartenbeck2019computational}.

Eq. \ref{eq:EFE_definition} can be used to calculate EFEs for second-stage actions. However, for first-stage actions, the level of optimality depends upon the EFEs of the final states and the transition probabilities. Therefore, for the first-stage actions, the EFE equation is given by: 

\begin{equation}
G(a_j) = p(s_B | s_A, a_j) \sum_{a_2 \in \mathcal{A}_B} G(a_2) \;+\; p(s_C | s_A, a_j) \sum_{a_2 \in \mathcal{A}_C} G(a_2)
\label{eq:first_stage_EFE} % Optional: add a label for referencing
\end{equation}
Where $\mathcal{A}_B$ and $\mathcal{A}_C$
are the sets of available actions in the second-stage states $s_{B}$ and $s_C$ respectively, and $G(a_2)$ are the second-stage EFEs given by Eq. \ref{eq:EFE_definition}.

Previous studies have reported a tendency for participants to repeat initial-stage actions, regardless of the outcome history \cite{gijsen2022active}. Within AIF, this behaviour is formalised through a habit term $E$. In the model of [4], $E$ is parameterised by $\kappa$, which modulates the extent of an agent’s reliance on habits (see Eq. \ref{eq:habit}, Appendix \ref{apx_aif_learning_rules}).

At the first stage, an agent will select an action that minimizes the EFE corrected by the habitual bias $G_t^\text{net}(s_1,a_1) = G_t(s_1,a_1) +E(a_{1}) $. At the second stage, no habitual bias is assumed, and therefore $G_t^\text{net}(s_2,a_2) = G_t(s_2,a_2)$. 

For the original AIF model \cite{gijsen2022active}, the choice probabilities of an agent at trial $t$, stage $p = \{1,2\}$ and state $s$, will be given by the softmax distribution of the corresponding net EFEs parametrized by an inverse temperature parameter, $\gamma_p$ (one for each stage). However, for the AIF-DDM the choice will be determined by a drift-diffusion process, with a non-decision time $t_{\text{nd}}$, boundary separation $a_{bs}$ and a drift rate $v_{p,s,t}$, that depends on the difference in $G^{\text{net}}$ of the two available actions such that:

\begin{equation}
v_{p,s,t} = v^{\text{mod}}_{p}[G^{\text{net}}_t(s,a) - G^{\text{net}}_t(s,a')]\, , \:\: a,a'\in \mathcal{A}_s
\label{eq:drift_rate_aif} 
\end{equation}

Where $\mathcal{A}_s$ is the set of available actions at state $s$, and  $v^{\text{mod}}_{p}$ is a free parameter that regulates the sensitivity to $G^{\text{net}}$ differences (which can also be interpreted as the agent's information-processing speed). Following the original AIF model \cite{gijsen2022active}, we fit a separate $v^{\text{mod}}_{p}$ parameter for each stage, an approach analogous to the use of stage-specific inverse temperatures.

In the classical DDM, a starting-point bias parameter $z$ is often included. However, in the two-step task, the symbols representing the different actions were randomly displayed either on the left or the right of the screen \cite{feher2020humans}. Therefore, we set $z = 0.5$ (effectively cancelling the bias in the drift-diffusion process).

After completing every trial, the agent updates its beliefs about the second-stage outcome probabilities. In \cite{gijsen2022active}, the updating rules for the prior distribution over outcome probabilities (Eqs. \ref{eq:alpha_update} and \ref{eq:beta_update} in Appendix \ref{apx_aif_learning_rules}), depend on four free parameters: the learning rate $l$, the prior volatility, $v_{\text{PS}}$, which modulates the influence of surprise on the agent's beliefs, volatility of sampled actions $v_{\text{SD}}$, which modulates how beliefs over chosen actions \textit{decay}, or are forgotten, and the volatility of unsampled actions $v_{\text{UD}}$, similar to $v_{\text{SD}}$ but for unchosen actions.

The study by [3] tested four AIF variants distinguished by their learning rules: a No Unsampled-Decay model ($\text{AIF}_\text{NUD}$, $v_{\text{UD}}=0$); a No Sampled-Decay model ($\text{AIF}_\text{NSD}$, $v_{\text{SD}}=0$), a No Predictive Surprise model ($\text{AIF}_\text{NPS}$, $v_{\text{PS}}=0$) and a model including all the learning mechanisms ($\text{AIF}_\text{FULL}$). Although solid evidence was found favouring AIF over HRL, the best-fitting AIF variant could not be determined \cite{gijsen2022active}. Consequently, we fitted all four models in our analysis.

\subsection{Cognitive effort and active inference}
A recent formalisation of cognitive effort ($\xi$) based on AIF defines it as the KL divergence between context-sensitive beliefs about how to act $P_{G}(\pi)$ and context-insensitive prior beliefs, or habits $P_{E}(\pi)$ \cite{parr2023cognitive}:

\begin{equation} \label{eq:cognitive_effort_def}
\underbrace{\xi \triangleq D_{\text{KL}}[P_G(\pi) || P_E(\pi)]}_{\text{Effort}}
=
\underbrace{\mathbb{E}_{P_G}[\ln P_G(\pi)]}_{\text{Context sensitive}}
-
\underbrace{\mathbb{E}_{P_G}[\ln P_E(\pi)]}_{\text{Context insensitive}}
\end{equation}

\vspace{5pt} 
\begin{equation*}
P_E(\pi) = \text{Cat}(\sigma(-\mathbf{E}))
\qquad
P_G(\pi) = \text{Cat}(\sigma(-\mathbf{G} - \mathbf{E}))
\end{equation*}

Where $\mathbf{G}$ and $\mathbf{E}$ are
vectors comprising the context-sensitive EFEs and the context-insensitive priors (or habits) of the available actions, respectively. 

From Eq. \ref{eq:cognitive_effort_def}, two key predictions can be derived. First, high effort will be experienced when there is an incongruence between \textbf{G} and \textbf{E}. Second, when the elements of $\mathbf{G}$ are of similar magnitude to one another, cognitive effort is minimal regardless of $\mathbf{E}$.

For the first stage of the two-step task, the AIF-DDM aligns with the first prediction of Eq. \ref{eq:cognitive_effort_def}. For a choice between actions $a_A$ and $a_B$, with $\mathbf{G} = [G_{A}, G_{B}]$, and $\mathbf{E} = [E_{A}, E_{B}]$, RTs should increase as the magnitude of the drift rate decreases, which is proportional to  $|\Delta G^{\text{net}}| = |G_{A}+E_{A}-G_{B}-E_{B}|$. Therefore, higher RTs may occur when, for example $|G_{A}|\gg|G_{B}|$, $|E_{A}|\ll|E_{B}|$ and $|G_{A}|\approx |E_{B}|$. In other words, when there is an incongruence between \textbf{G} and \textbf{E}.

However, our model contradicts the second prediction of Eq. \ref{eq:cognitive_effort_def}, because for a constant \textbf{E},  RTs should increase as $|G_{A}|$ gets closer to $|G_{B}|$. Interestingly,  for the second stage, Eq. \ref{eq:cognitive_effort_def} predicts minimal effort if we assume no habits (i.e. $E_{A}= E_{B}=0$), in contrast to AIF-DDM, where effort can still be high if the context-sensitive EFEs are closely matched.

\subsection{Model fitting and comparison procedures}
We followed the same Maximum Likelihood Estimate (MLE) procedure as in \cite{gijsen2022active}, to fit the models' free parameters to the behavioural data \footnote{\textbf{Code} available at https://github.com/decide-ugent/aif-ddm}. For each participant and model, we found the parameter set with the highest likelihood using Scipy’s ’L-BFGS-B’ algorithm \cite{virtanen2020scipy}. This step was repeated 35 times for each participant, with different (uniformly) randomized initializations for all parameters. After completing all the runs, we selected the parameter set with the maximum likelihood and used its value for model comparison.

For the pure AIF and HRL models, we computed the likelihoods using the choice probability distributions of each model. For AIF-DDM and HRL-DDM however, the likelihoods were given by the Wiener's First-Passage Time Distribution (WFPT) provided by the HDDM Python package \cite{wiecki2013hddm}. Further details can be found in Appendix \ref{apx_fitting_procedure}.

For model comparison, we relied on the Statistical Parametric Mapping (SPM) \cite{SPM2022FirstImages} and Variational Bayesian Analysis (VBA) \cite{daunizeau2014vba} toolboxes, using MATLAB R2022b. We performed a group-level random-effect Bayesian model selection (BMS) procedure, which requires the log-model evidence (LME) of each model for each participant. We used two LME approximations: the Bayesian Information Criterion (BIC) and the Akaike's Information Criterion (AIC) scores \cite{penny2012comparing} (see Appendix \ref{apx_fitting_procedure}). Using the BMS procedure, we estimated four model comparison metrics: the Expected Posterior Probability (EPP); the Estimated Model Frequency (EMF), or the proportion of participants best fit by the model; the Exceedance Probability (EP), which quantifies the likelihood that the model is more frequent than competing models across participants; and the Protected Exceedance Probability (PEP), similar to EP but accounting for the possibility that apparent differences in model frequencies arise due to chance \cite{rigoux2014bayesian}.

The SPM toolbox allows comparing the performance of model \textit{families} (i.e., models that share a common feature). As in \cite{gijsen2022active}, we use this feature to compare the overall AIF performance (considering its four variants) to HRL. Since PEPs are unavailable for family-level inference, we only calculate EPs for these analyses.

\subsection{Model and parameter recovery procedures}
To perform the parameter recovery analysis, we simulated a dataset for each model using parameter values sampled from uniform
distributions with ranges equal to those used to fit the real dataset. Next, we fitted the synthetic dataset and obtained new (recovered) parameters. Finally, we checked if the original and recovered parameters were correlated by calculating Pearson's correlations. This process was repeated 23 times (equal to the number of participants in the real experiment) for each model.

For the model recovery analysis, each model was first used to simulate a full experimental dataset with the same parameter sets as in the parameter recovery analysis. We then fitted all models to each simulated dataset and calculated the proportion of times each model was identified as best-fitting. Results were summarized in a confusion matrix (Fig. \ref{fig:confusion_matrix}).

\section{Results}

\subsection{Model recovery analysis}

\begin{figure}
    \centering

    \includegraphics[width=0.9\textwidth]{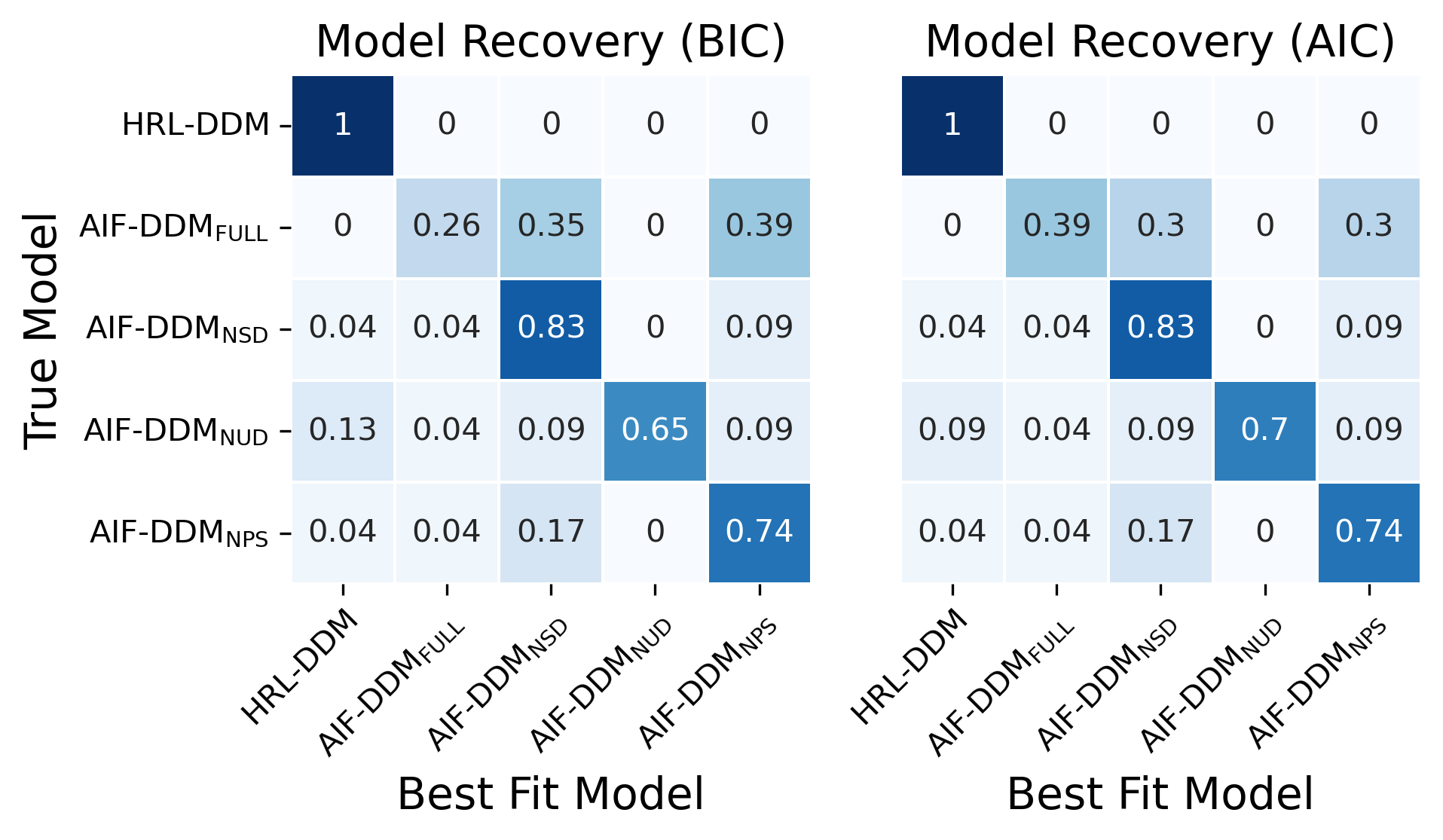}
    
    \caption{Model recovery results. Each cell contains the fraction of simulated experiments from a given true model, for which a corresponding model was classified as the best-fitting, according to the BIC score (left) or the AIC score (right).}
    \label{fig:confusion_matrix}
\end{figure}

Model recovery analysis indicates that $\text{AIF-DDM}_{\text{FULL}}$ is frequently misclassified as either $\text{AIF-DDM}_{\text{NSD}}$ or $\text{AIF-DDM}_{\text{NSD}}$ (Fig. \ref{fig:confusion_matrix}). The poor recovery likely stems from the full model encompassing all update rules present in the other two models \cite{gijsen2022active}. The rest of the models had a better but modest recovery (except HRL-DDM, which had a perfect recovery). Thus, combining AIF with a DDM  could not help substantially to differentiate between different learning variants as we hypothesised.
Overall, both AIC and BIC scores show similar results. Although the AIC score seems slightly more reliable since it could achieve a better recovery for both $\text{AIF-DDM}_{\text{FULL}}$ and $\text{AIF-DDM}_{\text{NUD}}$.

\vspace{-0.5cm}

\subsection{Model comparison results}

The BMS analysis revealed different results depending on the score used. According to the BIC score, $\text{AIF-DDM}_{\text{NSD}}$ and $\text{AIF-DDM}_{\text{NPS}}$ dominated moderately across participants with Expected Posterior Probabilities, $\text{EPP} = [0.39,0.31]$, Protected Exceedance Probabilities, $\text{PEP} = [0.63,0.29]$ and Estimated Model Frequencies, $\text{EMF} = [0.44,0.35]$, respectively (see Fig. \ref{fig:ddm_models_comparison}, first row). However, according to the AIC scores $\text{AIF-DDM}_{\text{FULL}}$ was the best-fitting model with $\text{EPP} = 0.57$, $\text{PEP} = 0.97$ and $\text{EMF} = 0.7$ (see Fig. \ref{fig:ddm_models_comparison}, second row). The disagreement between AIC and BIC scores was also reported for the model comparison between pure HRL and AIF models \cite{gijsen2022active}, and likely stems from the fact that BIC penalizes complexity more than AIC.

Even though the two BMS procedures produced conflicting results, we selected $\text{AIF-DDM}_{\text{FULL}}$ as the best-fitting model, and focus exclusively on its results in the subsequent sections. This decision was based on the model recovery analysis, which showed that the full model is frequently misclassified as either $\text{AIF-DDM}_{\text{NSD}}$ or $\text{AIF-DDM}_{\text{NPS}}$. Nonetheless, we acknowledge that the BIC-based comparison favoured $\text{AIF-DDM}_{\text{NSD}}$ and $\text{AIF-DDM}_{\text{NPS}}$, so we replicated our analysis for these models and show their results in Appendix \ref{apx_model_comparison}. We excluded $\text{AIF-DDM}_{\text{NUD}}$ from further analysis as it performed poorly in both BMS procedures.
Finally, the model family selection analysis revealed similar results to those reported in the study that compared pure AIF and HRL \cite{gijsen2022active}. We found that the AIF-DDM family outperformed the HRL-DDM model according to both BIC (EP = 1, EMF = 0.9, EPP = 0.88) and AIC (EP = 1, EMF = 0.92, EPP = 0.9) scores.

\subsection{Parameter recovery analysis}

Integrating a DDM with AIF substantially improved the recovery of the parameters shared between AIF-DDM and pure AIF models (see Appendix \ref{apx:parameter_recovery_results}, Table \ref{apx:tabs:parameter_recovery_tables}). For the pure AIF models, parameter recovery was far from perfect. For example, for $\text{AIF}_{\text{FULL}}$, only $v_{\text{DU}}$ had a Pearson's correlation greater than .6 between its ground-truth and recovered values. In contrast, for the AIF-DDM models, $v_{\text{DU}}$, $v_{\text{DS}}$, $v_{\text{PS}}$, $a_{\text{bs}}$ and $t_{\text{nd}}$ showed excellent recovery ($r>.90$, $p<.001$), $l$ and $\kappa$ were well recovered ($r>.84$, $p<.001$) and $p_{r}$ and $\lambda$ had a moderate recovery ($r>.67$, $p<.001$). The drift rate parameters $v_{1}^{\text{mod}}$ and $v_{2}^{\text{mod}}$  had moderate-to-poor recovery, depending on the specific model. Nonetheless, their recovery was still superior to that of the inverse temperature parameters ($\gamma_1$ and $\gamma_2$) from the pure AIF model.

\subsection{Expected free energy discriminability affects second stage, but not first stage reaction times} \label{sec:simulation_analysis}

To evaluate the models' goodness-of-fit and predictive accuracy, we compared model-simulated behaviour on the two-step task to the observed behaviour of participants. This was done using a decile-binned analysis based on the absolute difference in the net EFE between the two actions, $|\Delta G^{\text{net}}|$. For each AIF-DDM model (except $\text{AIF-DDM}_{\text{NUD}}$), we first computed the trial-by-trial $|\Delta G^{\text{net}}|$ for each participant for both stages, based on their best-fit model parameters, choice history, and the experienced sequence of rewards and transitions. From the resulting distribution of $|\Delta G^{\text{net}}|$ values, for each participant and stage, we identified the decile boundaries ($10^{\text{th}}$-$90^{\text{th}}$ percentiles).
We then binned each trial's observed RT and choice into one of ten decile bins according to its $|\Delta G^{\text{net}}|$ value. For example, a trial with a $|\Delta G^{\text{net}}|$ value below the $10^{\text{th}}$ percentile was placed in the first bin ($0^{\text{th}}$-$10^{\text{th}}$). Within each bin, we calculated two metrics for each participant: (1) the mean RT, and (2) the probability of choosing the action with the lower net EFE, $P(\text{choose}\;G^{\text{net}}_{\text{min}})$. These participant-level metrics were then averaged across participants for each decile bin. Finally, for each trial stage, we generated 100 simulations to estimate the model's predicted RT and choice probability. These simulated metrics were then binned and averaged in the same way as the observed data.

The simulation analysis results were qualitatively similar across all models, therefore we only display the predictions of the $\text{AIF-DDM}_{\text{FULL}}$ model as an example in Fig. \ref{fig:simulation_analysis_results_AI_ddm0}. Equivalent plots for the remaining models are presented in the Appendix \ref{apx:rest_models_simulation_analysis}. In both stages, participants were more likely to select the action with the minimum net EFE as the value of $|\Delta G^{\text{net}}|$ increased. This trend was well-captured by all the AIF-DDM models (see Appendix \ref{apx:rest_models_simulation_analysis}).

The AIF-DDM models also predict that the mean RT should decrease as a function of EFE discriminability (i.e., as $|\Delta G^{\text{net}}|$ increases) for both stages. This effect can be observed for the second stage; however, it is almost negligible for the first one (Fig. \ref{fig:simulation_analysis_results_AI_ddm0}, bottom left).

\begin{figure}[ht!]
    \centering
    \captionsetup{skip=10pt}

    \includegraphics[width=1\textwidth]{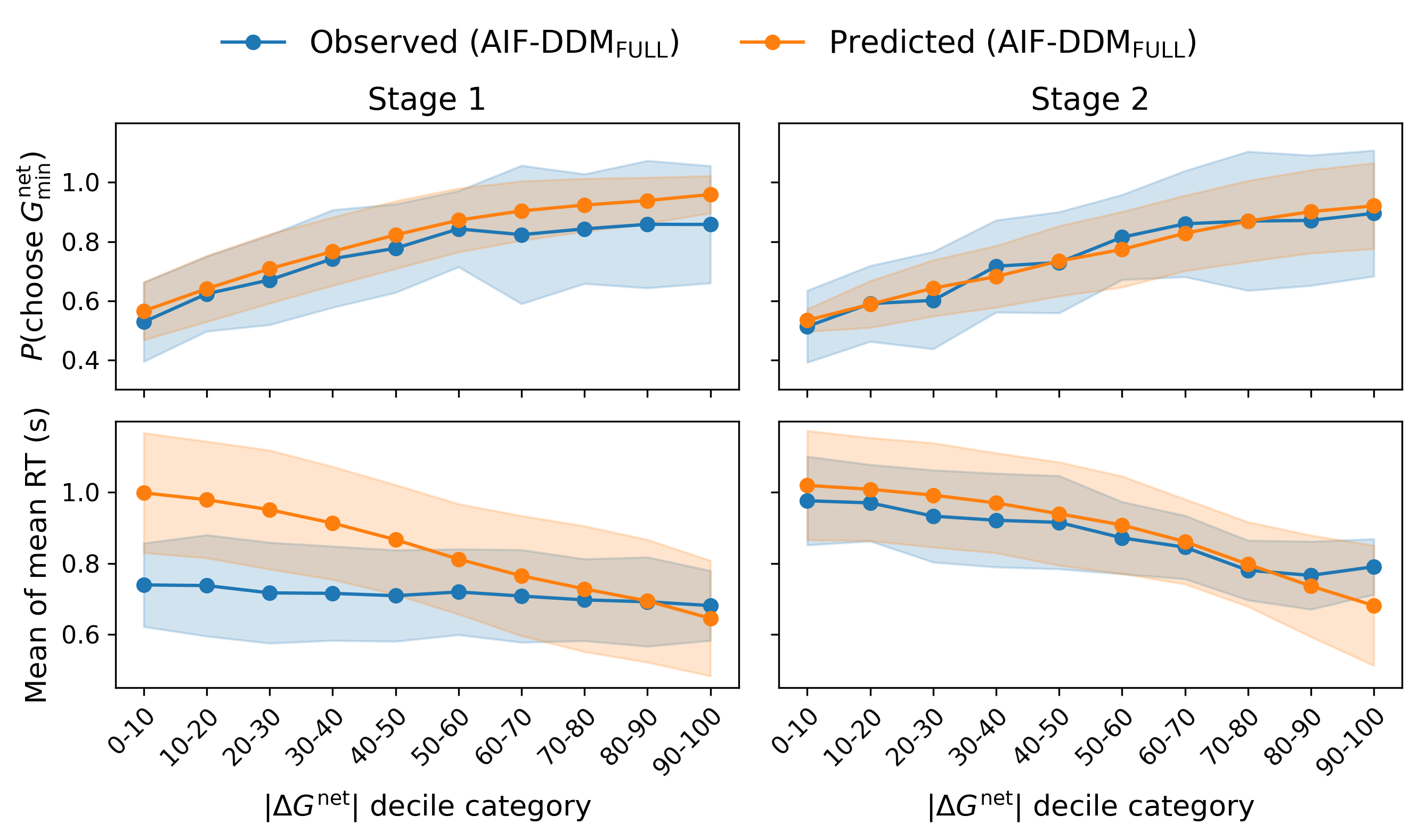}
    
    \caption{
    %Comparison of observed (blue) and model-predicted (orange) behaviour for choice probability (top) and reaction time (bottom). Data is binned by deciles of the difference in net Expected Free Energy ($|\Delta G^{\text{net}}|$). Lines represent the mean across participants; shaded areas are ±1 standard error of the mean. 
    Comparison of observed (blue) and model-predicted (orange) behaviour. Lines show the mean choice probability (top) and mean reaction time (bottom) across participants, binned by net Expected Free Energy ($|\Delta G^{\text{net}}|$) deciles. Shaded areas are ±1 standard error of the mean.}  
    \label{fig:simulation_analysis_results_AI_ddm0}
\end{figure}

\vspace{-0.05cm}

Multiple explanations may account for the mismatch in first-stage RTs. The first concerns the 2-second deadline. As described in Section 2.1, all missed trials were excluded from the analysis, effectively truncating the observed RT distribution. In contrast, the AIF-DDM models do not impose this time restriction, resulting in higher predicted RT means (see Appendix \ref{apx:qq_plots} for further details). However, this effect alone is unlikely to fully explain the misfit, since the second stage had the same deadline, yet a far less pronounced mismatch. 

\begin{figure}
    \centering

    \includegraphics[width=0.91\textwidth]{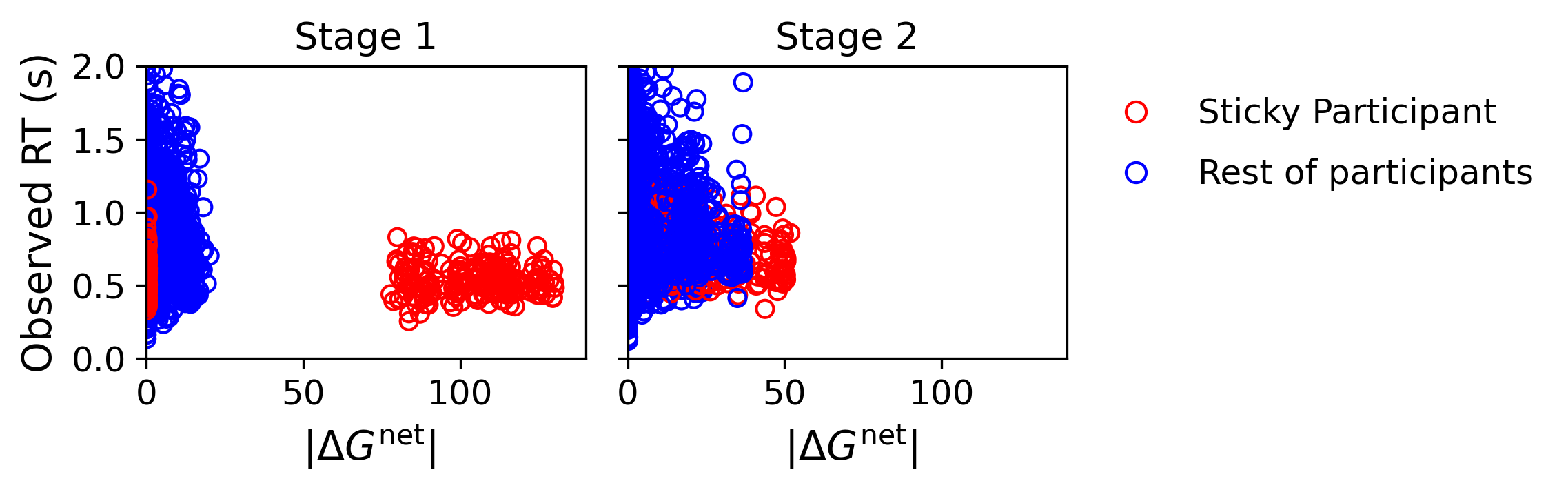}
    
    \caption{Observed reaction time (RT) vs. $|\Delta G^{\text{net}}|$ for the $\text{AIF-DDM}_\text{FULL}$ model, separated by task stage. Each circle represents a single trial. Trials from the 'sticky' participant, identified by the highest fitted $\kappa$ value, are shown in red. Trials from all other participants are shown in blue. For the sticky participant, notice the cluster of first-stage trials with high $|\Delta G^{\text{net}}|$ values compared to the second-stage trials that are more concentrated at low $|\Delta G^{\text{net}}|$ values.}
    \label{fig:delta_G_distr_AI_ddm1}
\end{figure}

A second possible explanation stems from the well-documented tendency for participants to repeat their previous first-stage choice \cite{gijsen2022active}. In both AIF(-DDM) and HRL(-DDM), a habit term is included only in the first-stage equations. It is therefore plausible that, for most first-stage choices, the habit strength of the previously selected option is sufficient to produce fast responses—even in trials falling into the lowest |$\Delta G^{\text{net}}$| deciles. In contrast, with the lack of a habit term in the second-stage equations, |$\Delta G^{\text{net}}$| values may be smaller or more variable. This effect was particularly notable for the participant with the largest fitted $\kappa$ value (where $\kappa$ regulates the tendency to repeat the previous first-stage choice). However, it was not observed for the rest of the participants (see Fig. \ref{fig:delta_G_distr_AI_ddm1}).

%Indeed, for both AIF(-DDM) and HRL(-DDM), a habit term is only included in the first-stage equations. Thus, it is conceivable that for most first-stage choices, the habit term of the previously chosen option is often large enough to trigger short responses, even for trials in the first |$\Delta G^{\text{net}}$| decile categories (Fig. \ref{fig:simulation_analysis_results_AI_ddm0}). In contrast, with the lack of a habit term in the second-stage equations, |$\Delta G^{\text{net}}$| values may be smaller or at least more variable. This effect was particularly notable for the participant with the largest fitted $\kappa$ value (where $\kappa$ regulates the tendency to repeat the previous first-stage choice). However, it was not observed for the rest of the participants (see Fig. \ref{fig:delta_G_distr_AI_ddm1}).

%In both AIF(-DDM) and HRL(-DDM), a habit term is included only in the first-stage equations. It is therefore plausible that, for most first-stage choices, the habit strength of the previously selected option is sufficient to produce fast responses—even in trials falling into the lowest |∆Gnet| deciles (see Fig. 3). In contrast, because the second-stage equations lack a habit term, |∆Gnet| values may be lower or more variable. This effect was particularly evident in the participant with the highest fitted κ value (where κ modulates the tendency to repeat the previous first-stage choice), but was not observed in the remaining participants (see Fig. 4).

A third potential explanation is that a substantial portion of the cognitive effort in the initial stage may arise from mental simulation or partial exploration of the decision tree, a process not accounted for by the assumption that RTs depend solely on |$\Delta G^{\text{net}}$|. Nevertheless, the limited complexity of the task (involving only four possible branches) constrains the extent of such planning.

\vspace{-0.6cm}

\subsection{The lack of net EFE discriminability effect on first-stage reaction times might be explained by the experimental design}
A final potential explanation for the lack of |$\Delta G^{\text{net}}$| effect on first-stage RTs is the experimental design. In the trial sequence of the two-step task, the elapsed time from the moment participants make the second-stage choice of the previous trial until they are presented with the first-stage options of the next trial is variable and long (3.2-3.8s). This time interval could allow participants to engage in "pre-thinking" about their upcoming first-stage choice before the options are presented. As a result, the recorded first-stage RT may only capture the final execution of an already made decision, rather than the full deliberation process. In contrast, the interval between the first- and second-stage choices is fixed and shorter (2s), leaving less room for pre-thinking and making the second-stage RT a more reliable measure of cognitive effort.

The difference in inter-choice intervals could also explain why mean first-stage RTs are shorter than mean second-stage RTs (see Fig.  \ref{fig:simulation_analysis_results_AI_ddm0}), which may seem counterintuitive since, according to AIF, participants in the first stage should engage in planning, a more costly cognitive process than the one-shot second-stage decision.

\section{Discussion}

In this study, we introduced a novel AIF-DDM and evaluated its predictions in the two-step task. We discussed the difference between the model predictions and a recently developed information-theoretic formalism of cognitive effort based on AIF (Eq. \ref{eq:cognitive_effort_def}, \cite{parr2023cognitive}). 

Our model predicts that both the need to overcome a pre-existing habit and the difficulty of discriminating between options with similar subjective values (EFEs) increase cognitive effort during the first stage of the two-step task. We argue that for the first stage, both effects can be regarded as a single competition process between the 'habit-corrected' or net EFEs of the available options. For the second stage, habits are assumed to play no role, and therefore, the model predicts that cognitive effort only depends on the discriminability between pure context-sensitive EFEs.

We found empirical evidence that EFE discriminability affects second-stage RTs of the two-step task as predicted by the AIF-DDM model -- an effect that is not captured by the information-theoretic formalism. However, we found that the discriminability between the habit-corrected EFEs had a negligible effect on the first-stage RTs, contrary to the AIF-DDM predictions. We argue that the latter observation could be a result of the long inter-trial intervals that may allow participants to "pre-think" their choice before being presented with the options. Thus, first-stage RTs may only reflect motor execution, not the preceding cognitive effort.

Lastly, we found that integrating a DDM with AIF did not help to distinguish between different learning mechanisms proposed by \cite{gijsen2022active}. However, the DDM integration improved parameter recovery compared to pure AIF. The latter result was expected, since the same effect has been reported for HRL in the two-step task \cite{shahar2019improving}. Nonetheless, the result might be relevant for computational psychiatry studies interested in finding the link between certain AIF parameters and specific pathologies.

Our study has multiple limitations. Firstly, we focused on comparing the predictions of AIF-DDM with a single information-theoretic definition of cognitive effort. Our intention, however, is not to disprove this particular formalism, but to illustrate that a more complete theory of cognitive effort should include both the effect of subjective value competition and the violation of prior expectations. Secondly, we focused on studying how cognitive effort may result from the competition between the EFEs and prior beliefs of the different available options. We did not, however, address the costs associated with learning or with \textit{generating} EFEs, which are likely to differ between the two stages. Finally, cognitive effort is a complex phenomenon that may depend upon multiple factors, such as the objective cognitive demand of a task or the participant's information-processing speed. Some of these factors may be captured by different AIF-DDM parameters (e.g., $a_{\text{bs}}$ and $\kappa$ for impulsiveness and motivation, respectively); however, this analysis was beyond the scope of this paper.

A future study could evaluate whether the pre-thinking effect is responsible for the AIF-DDM misfit for the first-stage RTs by reducing the inter-trial intervals. This could be achieved by, for example, reducing the amount of time that the rewards are displayed. Moreover, the problem of ruling out missed trials could be mitigated by increasing the available time to make a choice, using a likelihood function that can take into account the probability of missing a deadline, or introducing collapsing thresholds in the DDM \cite{voskuilen2016comparing}. Implementing these changes could move us closer to a better understanding of cognitive effort in the two-step task.

\appendix

\section{Further details on Active Inference in the two-step task}\label{apx_aif_learning_rules}

In \cite{gijsen2022active}, the distribution over prior preferred observations is given by the following equation:

\begin{equation}
P(o_t | C) = \frac{1}{Z(\lambda)} e^{o_t \lambda} e^{ - (1 - o_t) \lambda}
\label{eq:preferred_obs} 
\end{equation}

Where $\lambda$ is a free parameter that regulates how much an agent prefers to realize prior preferences (in this case, getting a reward) over gaining new information. Thus, $\lambda$ regulates the exploration-exploitation trade-off.

As mentioned in Section \ref{sec:aif_ddm}, agents are assumed to tend to repeat first-stage actions independent of the observed outcomes. This behaviour is modelled in AIF by introducing a \textit{habit} term. In \cite{gijsen2022active}, this quantity is described by:
\begin{equation}
E(a_j) = \frac{1}{Z(\kappa)} e^{\delta_{a_{t-1}, a_j} \kappa_e - (1 - \delta_{a_{t-1}, a_j}) \kappa}
\label{eq:habit} % Optional: add a label for referencing
\end{equation}

Where $\kappa$ is a precision parameter regulating the agent's reliance on habits.

After every trial, the AIF agent updates its beliefs about the second-stage outcome probabilities. Since these probabilities are mutually independent, the overall distribution equals the product of four beta distributions, each describing the believed outcome probability of its respective action and parameterized by its corresponding $\alpha$ and $\beta$.

\begin{equation}
\pi_t(\theta) = \prod_{s=1}^{2} \prod_{a=1}^{2} \mathcal{B}e(\alpha_{s,a,t}, \beta_{s,a,t})
\label{eq:theta2_prior} % Optional: add a label for referencing
\end{equation}

According to the learning rule implemented by \cite{gijsen2022active}, after observing the outcome of a chosen action, its corresponding $\alpha$ and $\beta$ parameters are updated following these rules:

\begin{equation}
\alpha_{s,a,t} = (1 - \chi_t)\alpha_{s,a,t-1} + \delta_{a_t, a} o_t l + (1-\nu_{SD})\alpha_{s,a,t-1}+\nu_{SD}\alpha_{0}
\label{eq:alpha_update} 
\end{equation}

\begin{equation}
\beta_{s,a,t} = (1 - \chi_t)\beta_{s,a,t-1} + \delta_{a_t, a}(1- o_t) l + (1-\nu_{SD})\beta_{s,a,t-1}+\nu_{SD}\beta_{0}
\label{eq:beta_update} 
\end{equation}

\begin{subequations}
\label{eq:chi_and_m} % Label for the whole group
\noindent
\begin{minipage}{0.48\textwidth}
\begin{equation}
\chi_{t} = \frac{m \;PS}{1+m\;PS}
\label{eq:chi_def_sub} % Label for (Na)
\end{equation}
\end{minipage}
\hfill
\begin{minipage}{0.48\textwidth}
\begin{equation}
m = \frac{\nu_{\text{PS}}}{1-\nu_{\text{PS}}}
\label{eq:m_def_sub}   % Label for (Nb)
\end{equation}
\end{minipage}
\end{subequations}

\indent

\vspace{0.1cm}

Where $\chi_{t}\in[0,1]$ is the surprise-modulated adaptation rate for trial $t$ which depends on the predictive surprise $PS = -\ln{p(o_t;\pi_{t}(\theta))}$, and the prior volatility parameter $\nu_{\text{PS}}\in[0,1]$ which modulates the influence of surprise on the agent's beliefs. In Eqs. \ref{eq:alpha_update} and \ref{eq:beta_update}, $l$ is the learning rate parameter, $\nu_{\text{SD}}$ is the volatility parameter of chosen actions and $\nu_{\text{SD}}$ modulates how over time $\alpha$ and $\beta$ decay towards their prior values $\alpha_{0}$ and $\beta_{0}$, respectively. 

Evidence from multi-armed bandit tasks suggests a bias in prior outcome probabilities. To model this effect, a free parameter, $p_{r}$, is fitted to each participant, which determines the initial (or uninformed) $\alpha$ and $\beta$ prior values, such that $\alpha_{0} = 2(1-p_r)$ and $\beta_{0}=2p_r$ \cite{gijsen2022active}. 

The decay effect of the chosen actions captured by $\nu_{\text{SD}}$, may also affect unchosen actions, and therefore, for these, the updating rules reduce to: 

\begin{equation}
\alpha_{s,a,t} = (1-\nu_{\text{UD}})\alpha_{s,a,t-1}+\nu_{\text{UD}}\alpha_{0}
\label{eq:unsamp_alpha_update} 
\end{equation}

\begin{equation}
\beta_{s,a,t} = (1-\nu_{UD})\beta_{s,a,t-1}+\nu_{UD}\beta_{0}
\label{eq:unsamp_beta_update} 
\end{equation}

Where $\nu_{\text{UD}}$ is the volatility of unchosen actions. 

The surprise update mechanism for chosen actions and the decay mechanisms for chosen and unchosen actions may not all be simultaneously operating \cite{gijsen2022active}. Therefore, four variants of the AIF models can be implemented, each with a different combination of update rules as pointed out in Section \ref{sec:aif_ddm}.

In addition to the outcome probabilities, agents must also learn the transition probabilities. As pointed by \cite{gijsen2022active}, before starting the experiment, participants completed an instruction phase as well as 50 practice trials and therefore, they were aware that the transition probabilities of initial-stage actions were mirrored and could either be $p(s_{B}|s_A, a_A) = p(s_C|s_A, a_B) = 0.7$ or $p(s_B|s_A, a_A) = p(s_C|s_A, a_B) = 0.3$ (with  $p(s_B|s_A,a_A)= 1 - p(s_C|s_A, a_A)$  and  $p(s_B|s_A, a_B) = 1 - p(s_C|s_A, a_B)$, respectively). In \cite{gijsen2022active}, this transition learning is modelled by having agents count transitions, and on each trial, choosing the most likely transition structure based on the observed frequencies.

\section{Hybrid reinforcement learning drift-diffusion model} \label{apx:hrl_ddm}
According to the HRL-DDM model, an agent performing the two-step task in state $s$, and trial $t$, will select an action $a$ that maximizes the expected discounted future reward (or state-action value), denoted by $Q_{t}(s, a)$. Traditionally, state-action values can be computed using either a model-free $Q_{t}^{\text{MF}}$ or a model-based strategy $Q_{t}^{\text{MB}}$.  

In the two-step task, after completing each trial, model-free state-action values are updated following the $\text{SARSA}(\lambda)$ algorithm. For the chosen second-stage action $a_{2}$ and the final state (either $s_{B}$ or $s_{c}$, depending on the transition), its corresponding state-action value is updated as follows:

\begin{equation}
Q^{\text{MF}}_{t+1}(s_{2},a_{2})= Q^{\text{MF}}_{t}(s_{2},a_{2}) + \alpha_{2}\delta_{2,t}
\label{eq:q_mf} 
\end{equation}

Where $\alpha_{2} \in[0,1]$ is the second-stage learning rate parameter and $\delta_{2,t}$, is the second-stage prediction error, defined as the difference between the observed outcome and its state-action value, such that $\delta_{2,t} = o_{t} - Q^{\text{MF}}_{t}(s_{2},a_{2})$

Since there is no outcome after choosing a first-stage action, and its optimality ultimately depends on the outcome observed in the second stage, the updating rule for the model-free state action value of the first-stage action will depend on the second-stage prediction error $\delta_{2,t}$ and will be given by the following equation:

\begin{equation}
Q^{\text{MF}}_{t+1}(s_{\text{A}},a_{1})= Q^{\text{MF}}_{t}(s_{\text{A}},a_{1}) + \alpha_{1}[\delta_{2,t}+\lambda\delta_{1,t}]
\label{eq:q_mf} 
\end{equation}

Where $\alpha_{1}\in[0,1]$ is the first-stage learning rate parameter, $\delta_{1,t}$, is the first-stage prediction error, defined as $\delta_{1,t} = Q^{\text{MF}}_{t}(s_{2},a_{2})-Q^{\text{MF}}_{t}(s_{\text{A}}, a_{1})$, and $\lambda \in[0,1]$ is the eligibility parameter, which determines how much the second-stage prediction error influences the first-stage state-action value. 

For the model-based strategy, agents are assumed to have a generative model of the task, which in this case means that they know (or learn) the state transition probabilities and use these to calculate state-action values. For the first-stage actions, the model-based values depend on the second-stage model-free values, such that:

\begin{equation}
Q_t^{\text{MB}}(s_{\text{A}}, a_{1}) = \sum_{s_2 \in \{s_{\text{B}},s_{\text{C}}\}} P(s_2 | s_\text{A}, a_{1}) \max_{a_2 \in \mathcal{A}_{s_2}} Q_t^{\text{MF}}(s_2, a_2)
\label{eq:q_mb} 
\end{equation}

In the first stage, the agent will follow a combination of model-free and model-based strategies, such that the total net expected reward of a given action $a_{j}$, at the first-stage state, $s_{A}$ is given by a linear combination of the trial's model-free $Q_{t}^{\text{MF}}$ and model-based $Q_{t}^{\text{MB}}$ expected rewards:

\begin{equation}
Q^{\text{net}}_{t}(s_{\text{A}},a_{j})= w Q_{t}^{\text{MB}}(s_{\text{A}},a_{j})+ (1 - w)Q_{t}^{MF}(s_{\text{A}},a_{j})+\rho \: \text{rep}_t(a_j)
\label{eq:q_net} 
\end{equation}

Where $w \in [0,1]$ is a weight parameter that regulates the relative influence of model-based and model-free strategies in the decision-making. The last term in the equation is added to model the tendency observed in participants to repeat the first-stage choice of the previous trial. The level of response stickiness is regulated by the parameter $\rho$, which multiplies the function $\text{rep}(a)$ -- equal to 1 if the first-stage action is the same as the one chosen in the previous trial and 0  otherwise.

At the second stage, the state-action values are computed using only a model-free strategy, and it is assumed that there is no tendency to repeat the previous action. Therefore, for the second stage, the net state-action value reduces to $Q^{\text{net}}_{t}(s_{2},a_{2})=Q^{\text{MF}}_{t}(s_{2},a_{2})$.

Once state-action values are computed, the HRL-DDM agent, will pick an action $a$ at trial $t$, state $s$ and stage $p=\{1,2\}$ through a drift-diffusion process with a drift rate given by the following equation:

\begin{equation}
v_{p,s,t} = v^{\text{mod}}_{p}[Q^{\text{net}}_t(s,a) - Q^{\text{net}}_t(s,a')]\, , \:\: a,a'\in \mathcal{A}_s
\label{eq:drift_rate_hrl} 
\end{equation}

Where $\mathcal{A}_s$ is the set of available actions at state $s$.

\section{Further details on the model fitting and comparison procedures}\label{apx_fitting_procedure}

For the two-step task, the MLE procedure consists of finding the participant's set of parameter values of the model $m$, $\hat{\theta}_m^{\mathrm{MLE}}$, that maximizes the likelihood of the first-stage and second-stage behavioural data ($d^{1}_{1:T}$ and $d^{2}_{1:T}$, respectively), given the parameters set, $p(d^{1}_{1:T},d^{2}_{1:T}|\theta_m,m)$, or equivalently, log-likelihood, $LL = \log \, p(d^{1}_{1:T},d^{2}_{1:T}|\theta_m,m)$. This quantity can be expressed as the sum of the individual trial LLs \cite{wilson2019ten}, such that:

\begin{equation}
\label{eq:LL} 
LL = \sum_{t=1}^{T} \log p(d^1_{t},d^2_t \mid d^1_{1:t-1}, d^2_{1:t-1}, \theta_m, m)
\end{equation}

To use a standard minimization procedure, instead of maximizing the LL, we minimize the negative of it (NLL). Moreover, we can express the LL of each trial's data as the product of the first-stage and second-stage data, giving the following expression for the NLL:

\begin{equation}
\label{eq:NLL} 
NLL = -\sum_{t=1}^{T} \log p(d^1_{t} \mid d^1_{1:t-1}, d^2_{1:t-1}, \theta_m, m) +\log p(d^2_t \mid d^1_{1:t}, d^2_{1:t-1}, \theta_m, m)
\end{equation}

Where $p(d^1_{t} \mid d^1_{1:t-1}, d^2_{1:t-1}, \theta_m, m)$ and $p(d^2_t \mid d^1_{1:t}, d^2_{1:t-1}, \theta_m, m)$
are the probabilities of each choice (or choice and RT) of the first and second stages, respectively,  given the parameters of the model and the information available up to that moment. 

For AIF and HRL models, $d^p_{t}$, corresponds to the choice made at stage $p$ and trial $t$, so Eq. \ref{eq:NLL} can be evaluated using the corresponding choice probability distributions of each model. For AIF-DDM and HRL-DDM, however, $d^p_{t}$, corresponds to the choice and RT tuple of stage $p$ and trial $t$. Hence, to compute the likelihood of each trial and phase, we use the WFPT distribution as implemented by the HDDM Python package. The WFPT distribution takes as input the parameters $a_{bs}$, $t_{nd}$, as well as the drift rates, which for HRL-DDM and AIF-DDM are calculated using  Eqs. \ref{eq:drift_rate_hrl} and \ref{eq:drift_rate_aif}, respectively.

To perform the BMS procedure for model comparison, we require an approximation for the LME of each model for each participant. Instead of calculating this quantity, we use two approximations; one based on the BIC score 
($LME \approx -BIC/2$) and another based on the AIC score 
($LME \approx -AIC/2$) \cite{penny2012comparing}, where the BIC and AIC scores are computed using the following equations:

\begin{equation}
    \mathrm{BIC} \coloneqq k \ln(n) - 2 \hat{LL}
\end{equation}

\begin{equation}
    \mathrm{AIC} \coloneqq 2k - 2 \hat{LL}
\end{equation}

Where $k$ is the number of free parameters in the model, $n$ is the number of trials, and  $\hat{LL}$ is the maximum $LL$ value, for a given participant and model.

\section{Model comparison plots}\label{apx_model_comparison}

\begin{figure}
    \centering

    \includegraphics[width=0.95\textwidth]{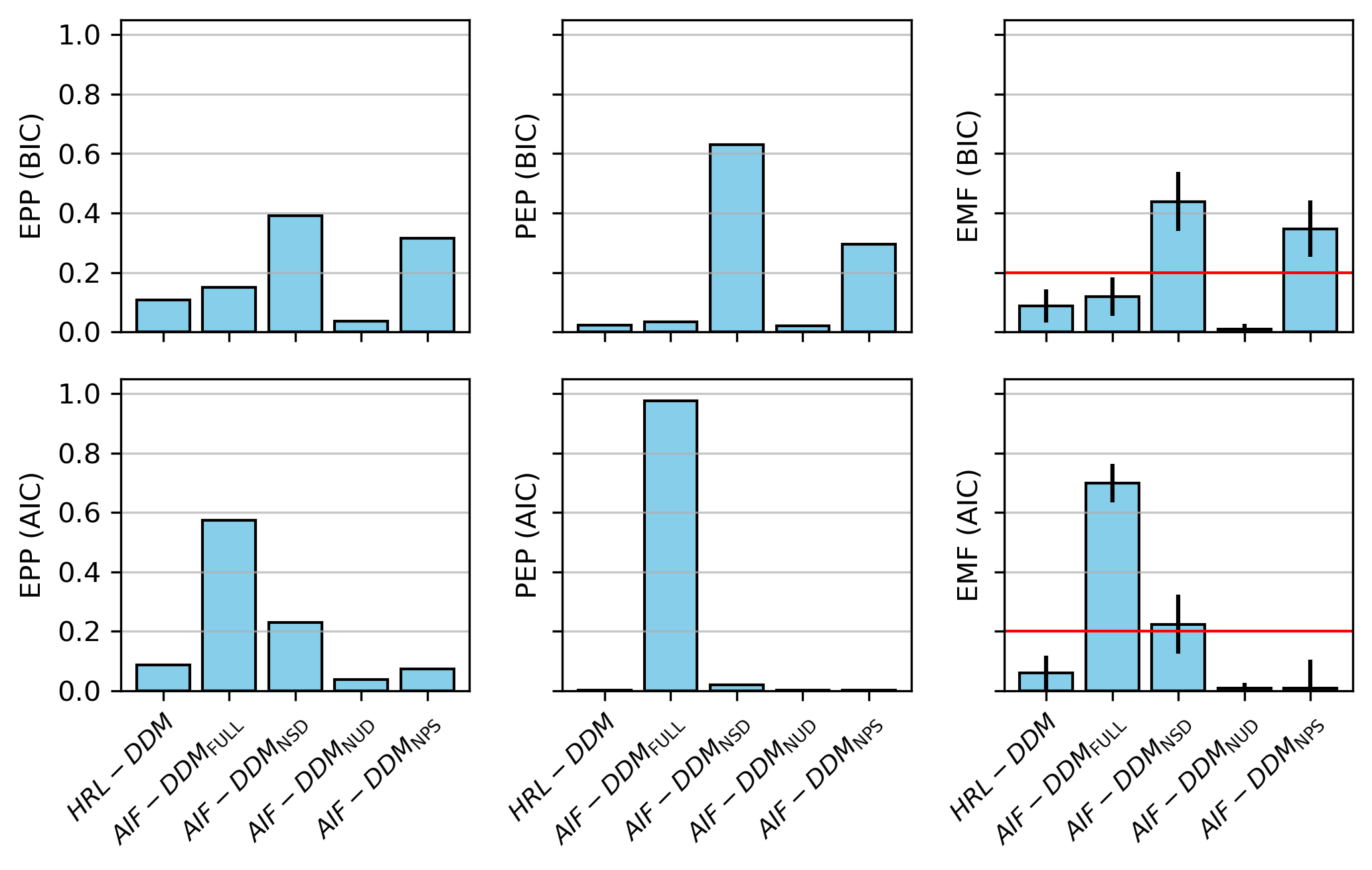}
    
    \caption{Bayesian model selection results comparing the five candidate models. The analysis was performed separately using log-model evidence approximations based on BIC (top row) and AIC (bottom row) scores. Each column represents a different metric for model comparison: (left) Expected Posterior Probabilities (EPP), (middle) Protected Exceedance Probabilities (PEP), and (right) Estimated Model Frequencies (EMF). In the EMF plots, the red horizontal line indicates the chance level (0.2), and error bars represent the standard deviation. The AIC-based analysis consistently favours the $\text{AIF-DDM}_{\text{FULL}}$ model, while the BIC-based results are more distributed across $\text{AIF-DDM}_{\text{NSD}}$, and $\text{AIF-DDM}_{\text{NPS}}$.}
    \label{fig:ddm_models_comparison}
\end{figure}
\clearpage
\section{Parameter recovery results}\label{apx:parameter_recovery_results}
%\vspace{2mm}
\begin{table}[ht!]
    \centering
    \caption{Parameter recovery results for the four AIF variants. The Pearson's correlation and p-values between original and recovered parameters for the pure AIF models are denoted by $r$ and $p$, respectively. For the AIF-DDM models, the Pearson's correlations and p-values are denoted by $r_{\text{ddm}}$ and $p_{\text{ddm}}$, respectively. } % Overall caption
    \label{apx:tabs:parameter_recovery_tables}

    \begin{subtable}{0.48\textwidth} % Adjust width as needed
        \centering
        \caption{$\text{AIF(-DDM)}_{\text{FULL}}$}
        \label{tab:full_aif}
        \small % Or \footnotesize, \small
        \setlength{\tabcolsep}{3pt} % Reduce column spacing
        \begin{tabular}{@{}lcccc@{}} % Using booktabs style, no vertical lines
        \toprule
        \textbf{Parameter} & \textbf{r} & \textbf{p} & \textbf{$\boldsymbol{r}_{\text{DDM}}$} & \textbf{$\boldsymbol{p}_{\text{DDM}}$}\\
        \midrule
        $l$ & .41 & .054 & .89  & $< .001$ \\
        $v_{\text{DU}}$ & .91 & $< .001$  & $>.99$ & $< .001$  \\
        $v_{\text{DS}}$ & .42 & $.043$ & .99 & $< .001$ \\
        $v_{\text{PS}}$ & .52 & .012 & .98 & $< .001$ \\
        $\lambda$ & .51 & .014 & .71 & $< .001$ \\
        $\kappa$ & .53 & .009 & .97 & $< .001$ \\
        $p_{r}$ & .28 & $.190$ & .93 & $< .001$\\
        $\gamma_{1}$ & .17 & .428 &  &  \\
        $\gamma_{2}$ & .38 & .073 &   &  \\
        $v_{1}^{\text{mod}}$ &  &  & .46 &  .027 \\
        $v_{2}^{\text{mod}}$ &  &  & .73 & $< .001$\\
        $a_{\text{bs}}$ &  &  & .98 & $< .001$\\
        $t_{\text{nd}}$ &  &  & >.99 & $< .001$ \\
        \bottomrule
        \end{tabular}
    \end{subtable}%
    \hfill % Pushes the next subtable to the right
    \begin{subtable}{0.48\textwidth}
        \centering
        \caption{$\text{AIF(-DDM)}_{\text{NSD}}$}
        \label{tab:nsd_aif}
        \small % Or \footnotesize, \small
        \setlength{\tabcolsep}{3pt} % Reduce column spacing
        \begin{tabular}{@{}lcccc@{}}
        \toprule
        \textbf{Parameter} & \textbf{r} & \textbf{p} & \textbf{$\boldsymbol{r}_{\text{DDM}}$} & \textbf{$\boldsymbol{p}_{\text{DDM}}$}\\
        \midrule
        $l$ & .62 & .002 & .96  & $< .001$ \\
        $v_{\text{DU}}$ & .90 & $< .001$  & .97 & $< .001$  \\
        $v_{\text{PS}}$ & .76 & $< .001$ & .98 & $< .001$ \\
        $\lambda$ & .49 & .017 & .68 & $< .001$ \\
        $\kappa$ & .68 & $< .001$ & .85 & $< .001$ \\
        $p_{r}$ & .83 & $< .001$ & .97 & $< .001$\\
        $\gamma_{1}$ & .10 & .644 &  &  \\
        $\gamma_{2}$ & .31 & .143 &   &  \\
        $v_{1}^{\text{mod}}$ &  &  & .68 &  $< .001$\\
        $v_{2}^{\text{mod}}$ &  &  & .59 & .003\\
        $a_{\text{bs}}$ &  &  & .99 & $< .001$\\
        $t_{\text{nd}}$ &  &  & >.99 & $< .001$ \\
        \bottomrule
        \end{tabular}
    \end{subtable}

    \vspace{1em} % Some vertical space between rows of subtables

    \begin{subtable}{0.48\textwidth}
        \centering
        \caption{$\text{AIF(-DDM)}_{\text{NUD}}$}
        \label{tab:nud_aif}
        \small % Or \footnotesize, \small
        \setlength{\tabcolsep}{3pt} % Reduce column spacing
        \begin{tabular}{@{}lcccc@{}}
        \toprule
        \textbf{Parameter} & \textbf{r} & \textbf{p} & \textbf{$\boldsymbol{r}_{\text{DDM}}$} & \textbf{$\boldsymbol{p}_{\text{DDM}}$}\\
        \midrule
        $l$ & .47 & .025 & .84  & $< .001$ \\
        $v_{\text{DS}}$ & .69 & $< .001$ & .93 & $< .001$ \\
        $v_{\text{PS}}$ & .57 & .005 & .94 & $< .001$ \\
        $\lambda$ & .66 & $< .001$ & .79 & $< .001$ \\
        $\kappa$ & .82 & $< .001$ & .93 & $< .001$ \\
        $p_{r}$ & .32 & $.131$ & .69 & $< .001$\\
        $\gamma_{1}$ & .06 & .798 &  &  \\
        $\gamma_{2}$ & .22 & .315 &   &  \\
        $v_{1}^{\text{mod}}$ &  &  & .55 &  .007 \\
        $v_{2}^{\text{mod}}$ &  &  & .60 & $ .003$\\
        $a_{\text{bs}}$ &  &  & .95 & $< .001$\\
        $t_{\text{nd}}$ &  &  & >.99 & $< .001$ \\
        \bottomrule
        \end{tabular}
    \end{subtable}%
    \hfill
    \begin{subtable}{0.48\textwidth}
        \centering
        \caption{$\text{AIF(-DDM)}_{\text{NSL}}$}
        \label{tab:nsl_aif}
        %\scriptsize % 
        \small
        \setlength{\tabcolsep}{3pt} % Reduce column spacing
        \begin{tabular}{@{}lcccc@{}}
        \toprule
        \textbf{Parameter} & \textbf{r} & \textbf{p} & \textbf{$\boldsymbol{r}_{\text{DDM}}$} & \textbf{$\boldsymbol{p}_{\text{DDM}}$}\\
        \midrule
        $l$ & .69 & $< .001$ & .93  & $< .001$ \\
        $v_{\text{DU}}$ & .94 & $< .001$ & .91 & $< .001$  \\
        $v_{\text{DS}}$ & .81 & $< .001$  & .97 & $< .001$ \\
        $\lambda$ & .56 & .005 & .67 & $< .001$ \\
        $\kappa$ & .68 & $< .001$ & .87 & $< .001$ \\
        $p_{r}$ & .94 & $< .001$ & .93 & $< .001$\\
        $\gamma_{1}$ & .18 & .413 &  &  \\
        $\gamma_{2}$ & .31 & .146 &   &  \\
        $v_{1}^{\text{mod}}$ &  &  & .25 &  .246 \\
        $v_{2}^{\text{mod}}$ &  &  & .58 & $ .004$\\ % Note: was < .004, not < .001
        $a_{\text{bs}}$ &  &  & $>.99$ & $< .001$\\
        $t_{\text{nd}}$ &  &  & $>.99$ & $< .001$ \\
        \bottomrule
        \end{tabular}
    \end{subtable}
    
\end{table}

\clearpage
\section{$\text{AIF-DDM}_{\text{NSD}}$ and $\text{AIF-DDM}_{\text{NPS}}$ simulation results}\label{apx:rest_models_simulation_analysis}
\begin{figure}
    \centering

    \includegraphics[width=0.90\textwidth]{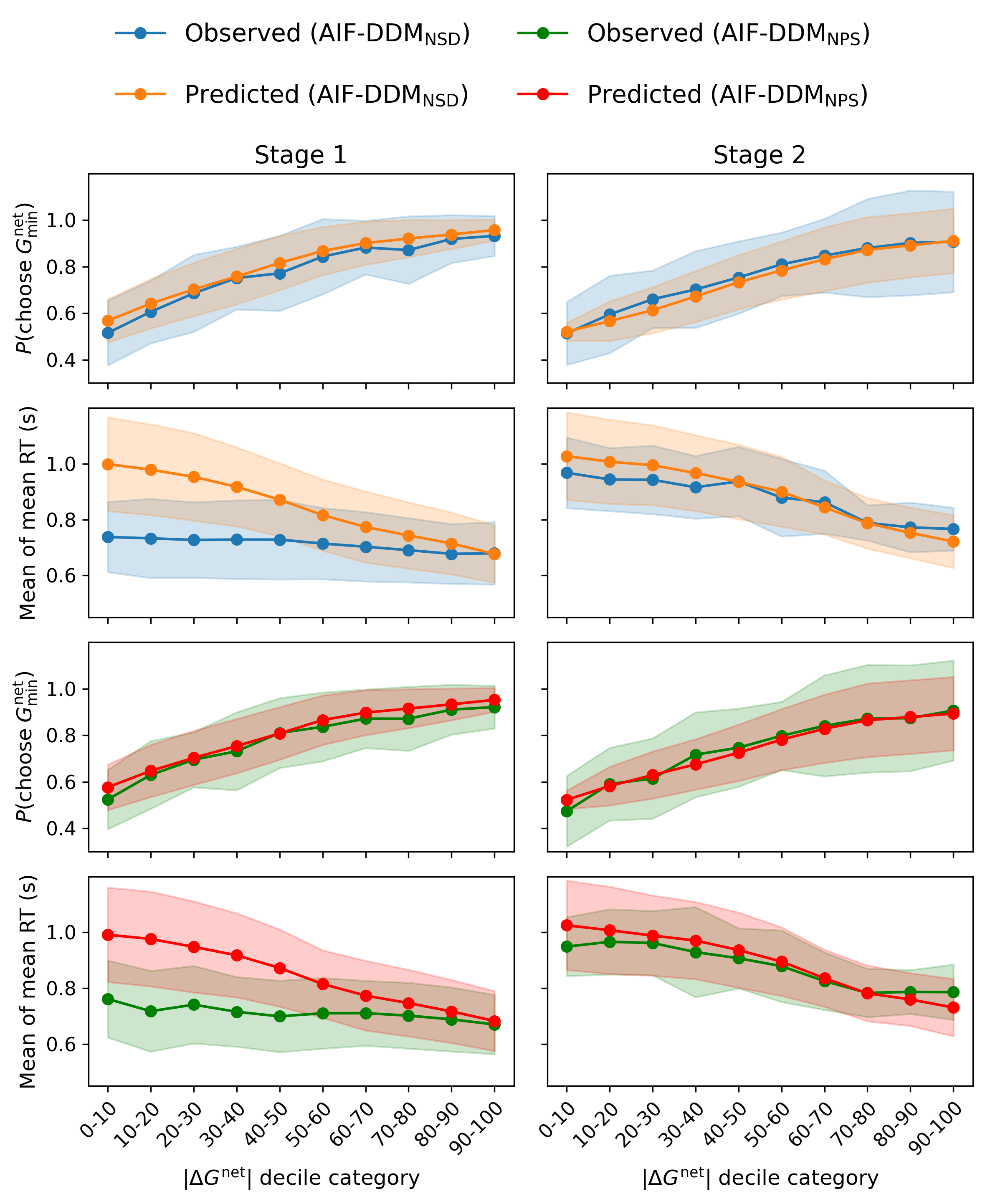}
    
    \caption{Comparison of observed and model-predicted behaviour for $\text{AIF-DDM}_{\text{NSD}}$ and $\text{AIF-DDM}_{\text{NPS}}$ models.}
    \label{fig:qq_plot}
\end{figure}
\clearpage
\section{Reaction time Q-Q analysis}\label{apx:qq_plots}

To further investigate the reaction time (RT) misfit mentioned in Section \ref{sec:simulation_analysis}, we generated Quantile-Quantile (Q-Q) plots comparing observed versus simulated within-participant RT deciles ($10^{th}-90^{th}$). Simulated RT deciles were produced for each participant by running 50 simulated experiments with 1000 trials each (with the same reward volatility as the real experiment) using the best-fitted parameters. As shown in Fig. \ref{fig:qq_plot}, the plots reveal that all models consistently overestimate the higher RT quantiles for both stages. This overestimation was expected, as the models' predictions are unconstrained, whereas the participants' observed RTs were truncated by a 2-second response deadline, with missed trials excluded from the analysis.

\begin{figure}
    \centering

    \includegraphics[width=1\textwidth]{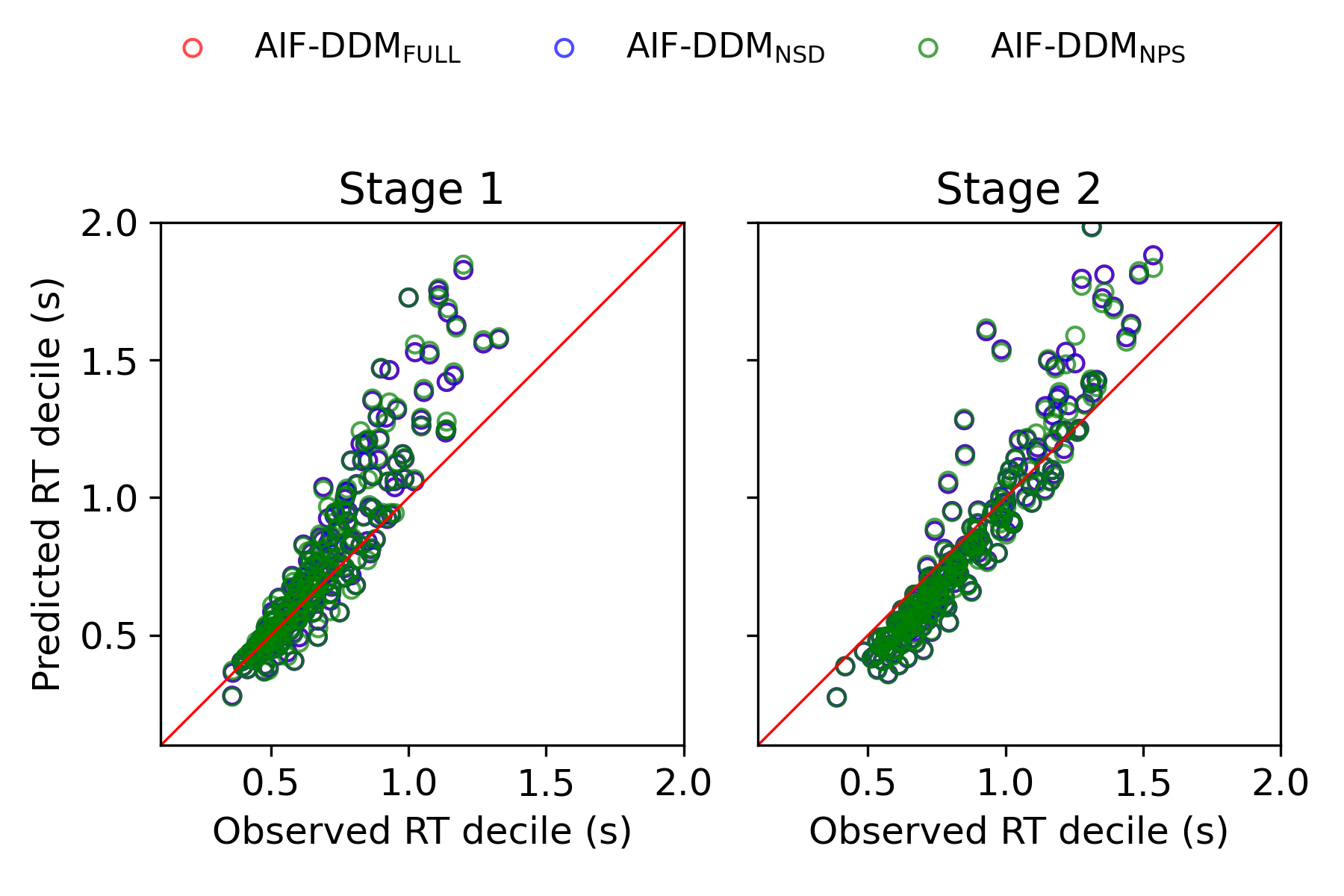}

    \caption{Systematic overestimation of higher reaction time (RT) deciles by the AIF-DDM models. The Q-Q plots compare observed versus predicted RT deciles for Stage 1 (left) and Stage 2 (right). Points lying above the identity line (red) indicate that the model's predicted RT is higher than the observed RT. This effect is most pronounced for the longest reaction times, where predictions deviate furthest from the identity line. Note that the purple markers indicate an overlap of red and blue points.}
    \label{fig:qq_plot}
\end{figure}

%\section{Relationship between Fitted kappa and mean deltaG}

%\begin{figure}
   % \centering
  %  \includegraphics[width=0.6\textwidth]{Figures/simulation_analysis/kappa_vs_deltaG_stage_AI_ddm1.png}
  %  \caption{Abstract representation of the two-step task}
  %  \label{fig:kappa_vs_deltaG_AI_ddm1}
%\end{figure}

\begin{credits}
\subsubsection{\ackname} 
%This work was supported by European Union’s Horizon 2020 FET research program under grant agreement No. 964464 (ChronoPilot).

This work was supported by European Union’s Horizon
2020 FET research program under grant agreement No.
964464 (ChronoPilot). We gratefully acknowledge the authors of \cite{feher2020humans} for making their dataset publicly available and the authors of \cite{gijsen2022active} for their open-source code.

\subsubsection{\discintname}
The authors have no competing interests to declare that are relevant to the content of this article.
\end{credits}
\bibliographystyle{splncs04}
\bibliography{mybibliography}


%\bibitem{ref_prior_preferred_obs}

\end{document}